\documentclass{aa}%[referee]
\usepackage[varg]{txfonts}
\usepackage{graphicx}

\usepackage{natbib}
\bibpunct{(}{)}{;}{a}{}{,}

\usepackage{color}

\begin{document}
  
\title{The close environment of high-mass X-ray binaries at high angular resolution}
\titlerunning{The close environment of high-mass X-ray binaries at high angular resolution I.}
\subtitle{I. VLTI/AMBER and VLTI/PIONIER near-infrared interferometric observations of \object{Vela~X-1}\thanks{Based on observations collected at the European Organization for Astronomical Research in the Southern Hemisphere, Chile. Programs 085.D-0029(A) and 088.D-0185(A).}}

\author{\'E.~Choquet\inst{1,2}\thanks{\email{choquet@stsci.edu}} 
	\and P.~Kervella\inst{1}
	\and J.-B.~Le~Bouquin\inst{3}
	\and A.~M\'erand\inst{4}
	\and J.-P.~Berger\inst{4}
	\and X.~Haubois\inst{5}
	\and G.~Perrin\inst{1}
	\and P.-O.~Petrucci\inst{3}
	\and B.~Lazareff\inst{3}
	\and J.-U.~Pott\inst{6}
}
\institute{LESIA, Observatoire de Paris, CNRS\,UMR\,8109, UPMC, Universit\'e Paris-Diderot, Paris Sciences et Lettres, 5 place Jules Janssen, 92195 Meudon, France
	\and Space Telescope Science Institute, 3700 San Martin Drive, Baltimore MD-21218, USA
	\and UJF-Grenoble 1/CNRS-INSU, Institut de Planétologie et d'Astrophysique de Grenoble (IPAG) UMR 5274, Grenoble, France
	\and European Southern Observatory, Alonzo de C\'ordova 3107, Casilla 19001, Santiago 19, Chile
	\and Instituto de Astronomia, Geof\`isica e Ci\^encias Atmosf\'ericas, Universidade de S\~ao Paulo, Rua do Mat\~ao 1226, Cidade Universit\'aria, S\~ao Paulo, SP 05508-900, Brazil
	\and Max-Planck-Institut für Astronomie, K\"onigstuhl 1, D-69117 Heidelberg, Germany
}

\date{Received 28 May 2013 / Accepted 4 November 2013}

\abstract{Recent improvements on the sensitivity and spectral resolution of X-ray observations have led to a better understanding of the properties of matter in the close vicinity of High Mass X-ray Binaries (HMXB) hosting a supergiant star and a compact object. However, the geometry and physical properties of their environment at larger scales (up to a few stellar radii) are currently only predicted by simulations but were never directly observed.}
{We aim at exploring the environment of \object{Vela~X-1} at a few stellar radii ($R_\star$) of the supergiant using spatially resolved observations in the near-infrared, and at studying its dynamical evolution along the 9-day orbital period of the system.}
{We observed \object{Vela~X-1} in 2010 and 2012 using near-infrared long baseline interferometry at the VLTI, respectively with the AMBER instrument in the K band (medium spectral resolution), and the PIONIER instrument in the H band (low spectral resolution). The PIONIER observations span through one orbital period to monitor possible evolutions in the geometry of the system.}
{We resolved a structure of $8\pm3~R_\star$ from the AMBER K-band observations, and $2.0\,_{-1.2}^{+0.7}~R_\star$ from the PIONIER  H-band data. From the closure phase observable, we found that the circumstellar environment of \object{Vela~X-1} is symmetrical in the near-infrared. We observed comparable interferometric measurements between the continuum and the spectral lines in the K band, meaning that both emissions originate from the same forming region. From the monitoring of the system over one period in the H band in 2012, we found the signal to be constant with the orbital phase within the error bars.}
{We propose three possible scenarios for this discrepancy between the two measurements: either there is a strong temperature gradient in the supergiant wind, leading to a hot component much more compact than the cool part of the wind observed in the K band, or we observed a diffuse shell in 2010, possibly triggered by an off-state in the accretion rate of the neutron star, that was dissolved in the interstellar medium in 2012 during our second observations or the structure observed in the H band was the stellar photosphere instead of the supergiant wind.}

\keywords{Techniques: interferometric, circumstellar matter, Stars: individual: \object{Vela~X-1}}

\maketitle

\section{Introduction\label{sec_intro}}
Supergiant High Mass X-ray Binaries (Sg-HMXB) are binary systems hosting a compact stellar remnant (black hole or neutron star) and a massive optical counterpart, usually an OB supergiant emitting a strong stellar wind. These systems are very bright in X-rays due to accretion of matter from the massive companion by the compact object that is deeply embedded in the supergiant wind in a very energy-efficient process \citep[see][for a review]{Chaty2011}. They are intensively observed in high energy spectral domains in order to better understand the processes at stake in accretion onto compact objects. In addition, as the X-ray luminosity is mostly related to the mass accretion rate in these systems, the compact object can be seen as a local probe of the density of matter evolving in the supergiant environment. This way, the clumpy nature of the wind of early-type supergiant has been analyzed and characterized by monitoring the variations in X-ray luminosity of Sg-HMXBs \citep{Walter2007}.

However, the way the compact object shapes this environment at a larger scale (up to a few tens of stellar radii) is still debated. Actually, several numerical simulations have been carried out, implementing different forces with varying intensities (depending on the physical and geometric properties of the systems). Some simulations predict that the supergiant wind could be focused by the gravitational field of the compact object and form a gaseous tail in its wake \citep{Hadrava2012}. Other simulations show that the velocity of the line-driven wind could be inhibited or even propelled back toward the supergiant by the X-ray heating and photoionization of the compact object, in case of close proximity in the binary system \citep{Hadrava2012,Krtivcka2012}. Finally, recent simulations revealed that in case of binary systems hosting a non-accreting pulsar, the colliding winds from both objects create shocked structures at 5 to 10 angular separations away from the stars \citep{Bosch-Ramon2012}.

To the best of our knowledge, such predicted structures in the environments of Sg-HMXBs have never been observed. Actually, these systems are usually at distances of a few kilo-parsec, with a typical size of a few tens of solar radii, making them very small targets with angular diameters of the order of a milli-arcsecond (mas). Only interferometric observations using baselines hundreds of meters long can spatially resolve the environment of these systems. We therefore carried out a pilot study on \object{Vela~X-1}, using the Very Large Telescope Interferometer (VLTI) in the near-infrared.

\object{Vela~X-1} is one of the most studied Sg-HMXB. The system is located at a distance of $d=1.9$~kpc from Earth \citep{Sadakane1985} and hosts a massive pulsar (identified as \object{4U~0900-40}) and a B0.5I supergiant (identified as \object{HD~77\,581} or \object{GP~Vel}) emitting a strong stellar wind. The mass and radius of the optical counterpart are estimated to be respectively $M_\star \sim32~\mathrm{M}_\sun$ and $R_\star \sim 24~\mathrm{R}_\sun$ \citep{Rawls2011}. Its wind has a terminal velocity of $V\sim1105~\mathrm{km.s}^{-1}$ \citep{Prinja1990} and a mass loss rate of $\dot M_\star \sim 1.5\times 10^{-6}~\mathrm{M}_\sun.\mathrm{yr}^{-1}$ \citep{Watanabe2006}. The pulsar has a spin period of 283~s \citep{McClintock1976}. Its mass $M_X\sim1.8~\mathrm{M}_\sun$ is known as one of the largest for a neutron star, and is regularly measured and refined through various methods \citep{Barziv2001,Quaintrell2003,Rawls2011,Koenigsberger2012}. Actually, its mass lies at the limit predicted by several theories on the still-unknown state of matter in neutron stars, and the precise estimation of the mass of the pulsar in \object{Vela X-1} can put constraints sufficient to rule out some of these theories. 
The pulsar is deeply embedded in the supergiant wind. It follows an eclipsing orbit with a period $P$ of $\sim9$~days \citep{Kreykenbohm2008}, and a projected semi-major axis $a\,\sin\,i\sim49~\mathrm{R}_\sun$ \citep{Rawls2011}, i.e. $\sim 2$\,R$_\star$. Accretion from this wind is the main source of X-ray emission in this system, with a luminosity of $L_X\sim 3.5 \times 10^{36}~\mathrm{erg.s}^{-1}$ \citep{Watanabe2006}.

From X-ray monitoring of the system luminosity, it seems that the pulsar evolves in a highly clumpy stellar wind, whose inhomogeneities in the density of matter produce important and irregular flares and off-states in X-rays \citep{Kreykenbohm2008,Furst2010,Doroshenko2011}. As usual for Sg-HMXBs, the characteristics of the environment at a larger scale are not clearly understood, and not directly observed up to now. Actually, UV \citep{Sadakane1985} and X-ray \citep{Watanabe2006} spectroscopic observations of \object{Vela~X-1} suggest the presence of a trailing wake behind the pulsar. In addition, the latter authors demonstrated that the supergiant dynamics is affected by the X-ray photoionization of the wind between the star and the pulsar, leading to a reduction of the velocity of the line-driven wind. This analysis has been recently confirmed through numerical simulations by \citet{Krtivcka2012}.

In order to spatially resolve the environment of \object{Vela~X-1} at a few stellar radii from the supergiant, we observed the system using near-infrared long baseline interferometry. We obtained two observations at VLTI in March 2010 and March 2012, with two different instruments, AMBER (K band) and PIONIER (H band). In Sect.~\ref{sec_obs} we describe the observations and instrumental setup. The data are analyzed in Sect.~\ref{sec_results}, and we discuss the results from the two data sets in Sect.~\ref{sec_discu}. Our conclusions are summarized in Sect.~\ref{sec_conc}.

\section{Observations and data reduction\label{sec_obs}}

\subsection{Instrumental setup}

Some useful characteristics of \object{Vela~X-1} are reminded in Table~\ref{tab_Vela_X-1}.  We got data at two different epochs on \object{Vela~X-1}, using two different infrared beam combiners at VLTI with two different sets of telescopes, in two different spectral bands. The first dataset was obtained in March 2010 with the AMBER instrument using the K spectral band (1.92--2.26~$\mu$m). The second dataset consists of four half nights spanning one orbital period of \object{Vela~X-1} in March 2012, and was acquired with the PIONIER instrument in the H band (1.5--1.8~$\mu$m). Overviews of the observations and of the telescope configurations are presented in Table~\ref{tab_date_obs} and \ref{tab_tel_conf} respectively, with the corresponding orbital phase of \object{Vela~X-1}. The orbital phase is computed using the ephemerids of \citet{Kreykenbohm2008}, with the X-ray mid-eclipse time as a reference for the 0-phase.

\begin{table}
  \caption{Characteristics of \object{Vela~X-1}.}\label{tab_Vela_X-1}
  \centering 
  \begin{tabular}{c c c}
  \hline\hline
  Parameter & Value & Ref. \\
\hline
\multicolumn{3}{c}{Binary system (\object{Vela~X-1}):}\\
\hline
	  $d$ (kpc) & $1.9 \pm 0.2$& 1\\ %Sadakane1985  il parle d\'ej\`a d'une queue derri\`ere la NS
	  $a\sin i$ (ls) & $113.98\pm0.13$ &2\\ %Bildsten1997
	  $P$ (d) & $8.964357\pm 0.000029$ &3\\ %\citet{kreykenbohm2008}
	  $T_{ecl}$ (MDJ) & $52974.227\pm 0.007$ &3\\ %\citet{kreykenbohm2008}
	  $e$ & $0.0898\pm0.0012$&2\\ %\citet{Bildsten1997}
	  $i$ (\degr) & $78.8 \pm 1.2$ &4\\% \citep{Rawls2011}

  \hline
\multicolumn{3}{c}{Pulsar (\object{4U~0900-40}):}\\
\hline
	    $L_X$ ($\mathrm{erg.s}^{-1}$)&$3.5 \times 10^{36}$&5\\ %\citet{Watanabe2006}   %Nagase1986
	    $M_X$ ($\mathrm{M}_\sun$)& $1.770 \pm 0.083$ & 4\\
  \hline
\multicolumn{3}{c}{Supergiant (\object{HD~77\,581}):}\\
\hline
	  $M_\star$ ($\mathrm{M}_\sun$)& $24.00 \pm 0.37$ & 4\\
	  $R_\star$ ($\mathrm{R}_\sun$)& $31.82 \pm 0.28$ & 4\\
	  $\dot M_\star $ ($\mathrm{M}_\sun .\mathrm{yr}^{-1}$) & $1.5-2 \times 10^{-6}$ & 5\\
	  $V$ ($\mathrm{km.s}^{-1}$)& $1105\pm 100$ & 6\\
  \hline 
  \end{tabular}
\tablebib{(1) \citet{Sadakane1985}; (2) \citet{Bildsten1997}; (3) \citet{Kreykenbohm2008}; (4)  \citet{Rawls2011}; (5) \citet{Watanabe2006}; (6) \citet{Prinja1990}.}
\end{table}

\begin{table}
  \caption{Telescope configurations.}\label{tab_tel_conf}
  \centering 
  \begin{tabular}{c c c c c}
  \hline\hline
  Instrument & Baseline & Length  & Position angle \\ 
	    &		&	(m)&(deg)\\
  \hline 
	  &UT3--UT4 & 62,5 & 110,8\\
  AMBER  &UT1--UT3 & 102,4 & 32,4 \\ 
&UT1--UT4 & 130,2 & 60,4 \\
\hline
&A1--I1 & 106,7 & 84,0\\
&A1--K0 & 129,0 & 63,9\\
PIONIER&A1--G1 & 80,0 & 107,9\\
&I1--K0 & 46,6 & 12,0\\
&I1--G1 & 46,6 & 40,0\\
&K0--G1 & 90,5 & 26,0\\
  \hline 
  \end{tabular}
\end{table}

\begin{table}
  \caption{Overview of the observations.}\label{tab_date_obs}
  \centering 
  \begin{tabular}{c c c c c}
  \hline\hline
  Date & Instrument & Band & MJD & Phase \\
  \hline 
  2010-03-26 & AMBER  & K &55281.135&0.342\\
  2012-03-25 & PIONIER& H &56012.129&0.887\\ 
  2012-03-27 & PIONIER& H &56014.106&0.107\\
  2012-03-29 & PIONIER& H &56016.088&0.328\\
  2012-03-31 & PIONIER& H &56018.100&0.553\\
  \hline 
  \end{tabular}
\end{table}

\subsubsection{AMBER observations}

AMBER is a three-telescope spectro-interferometer at VLTI, working in the J, H and K spectral bands (from 1 to 2.4~$\mu$m), and providing three different spectral resolutions ($R\sim30$, $R\sim1500$ and $R\sim12000$) \citep{Petrov2007}. The beams are spatially filtered using single-mode fibers to provide high precision interferometric measurements \citep{CoudeduForesto1998}. The interference pattern is acquired using multi-axial combination with a non-redundant spatial coding, simultaneously with a photometric channel for each beam so as to calibrate the coherent flux.

\begin{figure}
  \resizebox{\hsize}{!}{\includegraphics{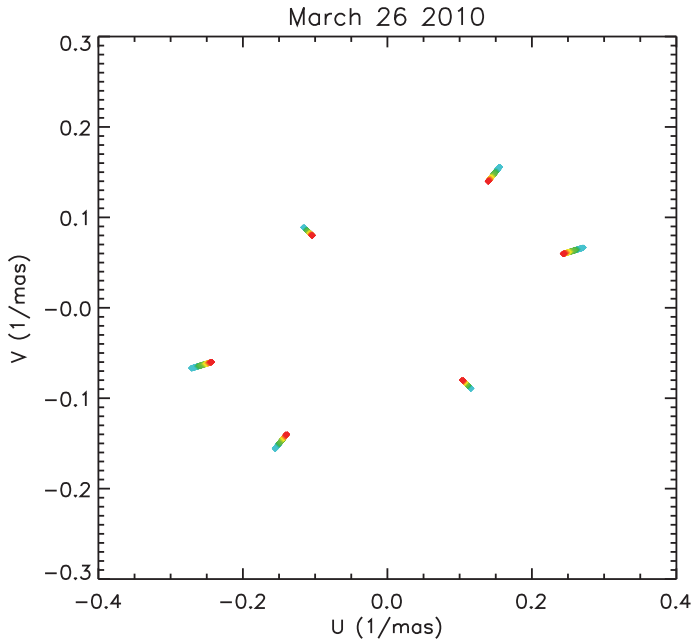}}
  \caption{Spatial frequencies of \object{Vela X-1} covered during the AMBER observations. The color of the symbol ranges with the wavelength of the observations, from blue for the shorter wavelength (2.03~$\mu$m) to red for the longest one (2.26~$\mu$m).}\label{fig_amber_uv}
\end{figure}

The AMBER observations of \object{Vela~X-1} were carried out using three 8~m Unit Telescopes (UTs) at VLTI (UT1, UT3, and UT4), with baseline lengths ranging from 63 to 130~m, using the K-band medium-resolution configuration of the instrument ($R\sim1500$). The (u,v) coverage obtained for this observation is presented in Fig.~\ref{fig_amber_uv}. The FINITO fringe tracker \citep{LeBouquin2008} was used for these observations to stabilize the fringes and improve the precision on the interferometric measurements.

\subsubsection{PIONIER observations}

PIONIER is a VLTI visitor instrument enabling the combination of four telescopes in the H band \citep{LeBouquin2011}. Each observation thus provides simultaneously six visibility and three independent closure phase measurements at different spatial frequencies, which enables the observer to constrain the spatial intensity distribution of the observed target. To achieve high precision interferometric measurements, the beams are spatially filtered using single-mode fibers, and combined using an integrated optics component \citep{Benisty2009}. Three different spectral resolutions can be used: broad band, $R~\sim15$, $R\sim35$. 

\begin{figure}
  \resizebox{\hsize}{!}{\includegraphics{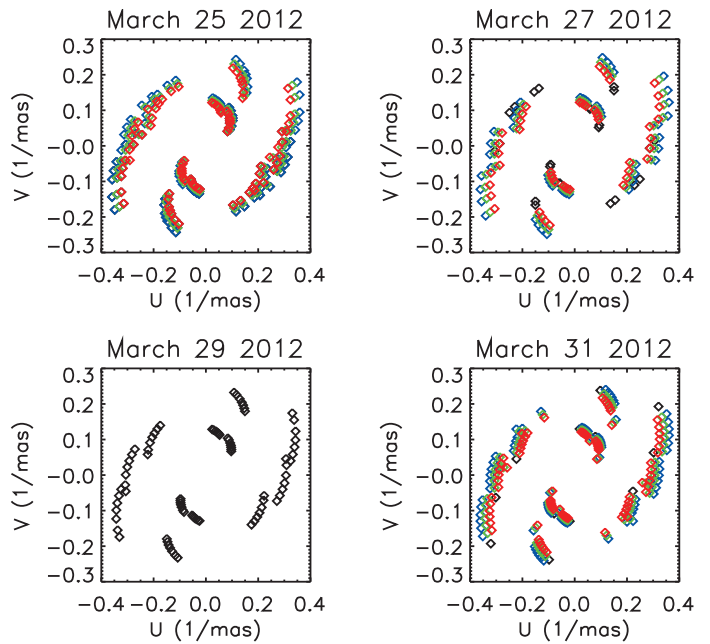}}
  \caption{Spatial frequencies of \object{Vela X-1} covered during the PIONIER observations, for each night. Black symbols stands for broad band measurements, colored symbols stands for low spectral resolution measurements (red, green and blue colors correspond to measurements at respectively 1.59~$\mu$m, 1.68~$\mu$m, and 1.76~$\mu$m).}\label{fig_pionier_uv}
\end{figure}

The PIONIER observations of \object{Vela~X-1} were carried out using the four relocatable 1.8~m Auxiliary Telescopes (ATs) at VLTI, on stations A1, G1, I1, K0, providing baseline lengths ranging from 47 to 129~m. Depending on the atmospheric conditions, either the broad band or the low resolution modes ($R\sim15$) were used. The log of the observations is presented in Table~\ref{tab_log_pionier}. The spatial frequencies covered by these observations are presented in Fig.~\ref{fig_pionier_uv}.

\begin{table}
  \caption{Log of the PIONIER observations.}\label{tab_log_pionier}
  \centering 
  \begin{tabular}{c c c c c}
  \hline\hline
  Date 		& Mode & \# Calibrated Points\tablefootmark{a} & Seeing \\ 
		&	&					& (\arcsec)\\
  \hline 
  2012-03-25 & Low res.& 7 & 0.5--0.9\\
  2012-03-27 & Low res. &4 &0.5--0.8\\
  2012-03-27 & Broad band &2 & 0.9--1.0\\
  2012-03-29 & Broad band & 7& 0.9--1.5\\
  2012-03-31 & Low res. & 7 & 0.7--1.1\\
  2012-03-31 & Broad band & 1 & 1.0\\
  \hline 
  \end{tabular}
  \tablefoot{\tablefoottext{a}{A calibrated point as defined by \citet{LeBouquin2011}}}
\end{table}

\subsection{Data reduction}
The calibrated interferometric measurements were obtained using the AMBER data reduction software \texttt{amdlib} package \citep[version 3.05;][]{Tatulli2007,Chelli2009} for the AMBER observations, and with the PIONIER data reduction software \texttt{pndrs}\footnote{Available at http://apps.jmmc.fr/~swmgr/pndrs} \citep[version 2.42;][]{LeBouquin2011} for the PIONIER observations. The stars used to calibrate the photometric and interferometric measurements are listed in Table~\ref{tab_calibrators}. 

The wavelength calibration of the AMBER data was obtained by fitting a part of the calibrated spectrum of \object{Vela~X-1} rich in telluric features with a model synthesized from spectroscopic parameters of atmospheric molecules using the HITRAN database, water vapor and atmospheric models, and optical fiber dispersion \citep{Merand2010}. For the AMBER analysis, measurements acquired at wavelengths lower than 2.03~$\mu$m are systematically discarded, since their quality is significantly worse than for longer wavelengths due to a strong atmospheric absorption.

The spectral calibration of the PIONIER data was obtained with the PIONIER data reduction software, as described in \citet{LeBouquin2011}. The authors estimate the accuracy of the calibrated effective wavelengths to be 2\%, which is sufficient for our analysis considering the 10\% maximal resolution of our observations.

\begin{table}
  \caption{Calibrators used to reduce the interferometric data.}\label{tab_calibrators}
  \centering 
  \begin{tabular}{c c c c}
  \hline\hline
  Name & Diameter & Ref. & Date(s) \\
      &	(mas)	  &		&\\
  \hline 
  HD~76304  & $0.749\pm0.010$ & 1& 2010-03-26 \\
  HD~74417  & $0.806\pm 0.010$ &1& 2012-03-25\\
  HD~75708  & $0.362\pm 0.026$ &2& 2012-03-27, 29, 31\\
  HD~76111  & $0.411\pm 0.029$ &2& 2012-03-25, 27, 29, 31\\
  HD~80934  & $0.467\pm 0.033$ &2& 2012-03-25, 27, 29, 31\\
  \hline 
  \end{tabular}
  \tablebib{(1)~\citet{Merand2005}; (2)~\citet{Lafrasse2010}.}
\end{table}

\section{Results\label{sec_results}}

\subsection{AMBER spectroscopic analysis\label{sec_AMBER_photom}}

We analyzed the spectrum of \object{Vela~X-1} obtained with AMBER and compared it to the 2~$\mu$m OB star atlas from \citet{Hanson1996}. Four spectral lines are clearly observed, with signals of respectively 4, 5, 3 and 2$\sigma$. They are referenced as three He~{I} transitions (at wavelengths 2.058, 2.113 and 2.162~$\mu$m) and the hydrogen Brackett-$\gamma$ (Br-$\gamma$) line at 2.166~$\mu$m. These recombination lines of Helium and Hydrogen are typical signatures of out-flowing gas close to OB supergiants \citep{Najarro1994}. The continuum emission can be related to the supergiant photosphere and to the free-free and free-bound emissions in the wind.

As a qualitative comparison with the spectra of B0.5 supergiants described in \citet{Hanson1996}, \object{Vela~X-1} shows similar spectral features as a B0.5 {Ib} star, with the same lines in emission and absorptions. This is consistent with previous classifications of \object{Vela~X-1}, which is described either as a B0.5 {Ia} \citep{Prinja2010}, a B0.5 {Iae} \citep{Krtivcka2012}, or a B0.5 {Ib} supergiant \citep{Dupree1980} in the literature. 

\begin{figure}
  \resizebox{\hsize}{!}{\includegraphics{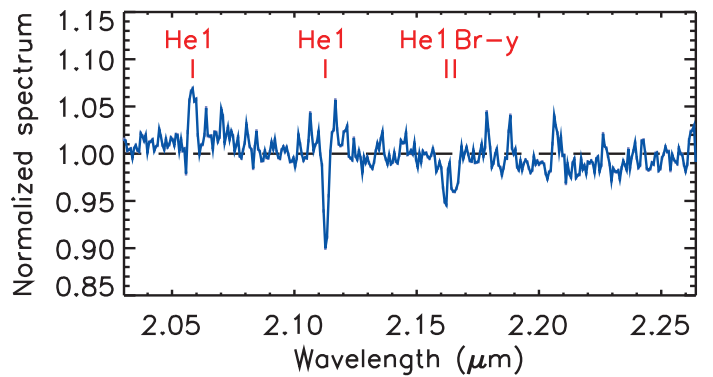}}
  \caption{Calibrated spectrum of \object{Vela~X-1} measured with AMBER. The error bars are about the size of the line (typically 2\%). The positions of four identified spectral features are marked by red solid lines. The dashed black line denotes the normalized value of the spectrum.}\label{fig_Amber_spectre}
\end{figure}

\subsection{AMBER interferometric analysis}
\begin{figure}
  \resizebox{\hsize}{!}{\includegraphics{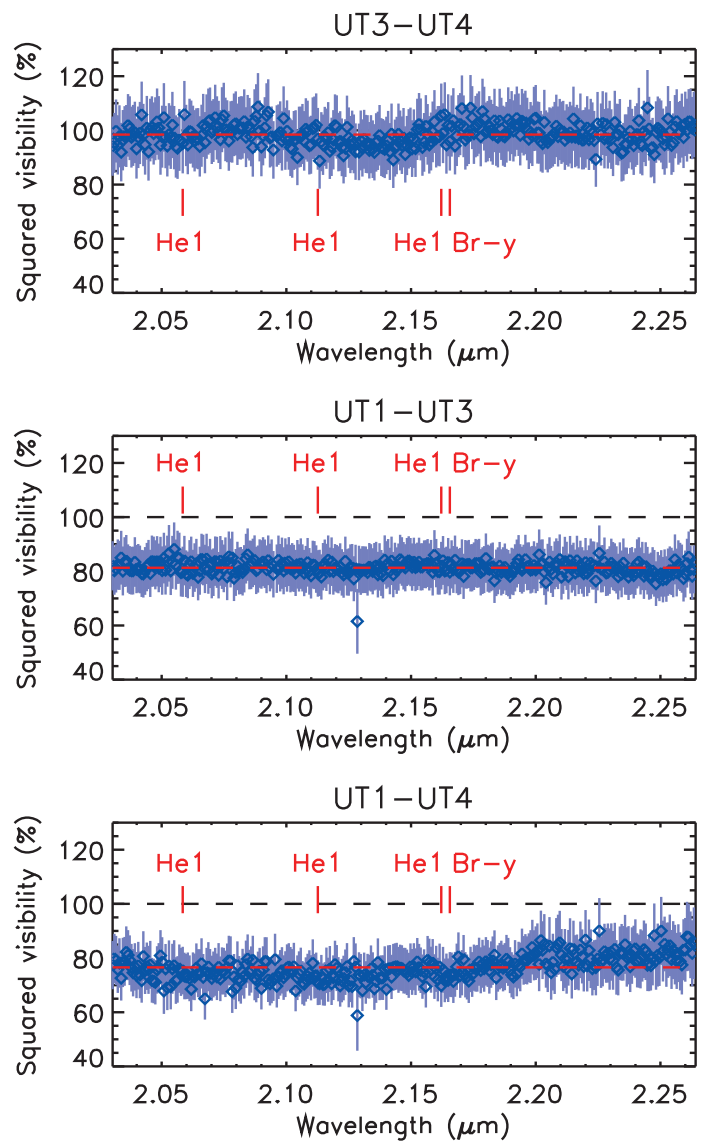}}
  \caption{Calibrated squared visibilities of \object{Vela X-1} obtained with AMBER on the three baselines. The dashed horizontal red lines mark the values in the continuum, and the solid vertical red lines mark the identified spectral features. The dashed black lines mark the squared visibility of an unresolved target.}\label{fig_amber_vis2}
\end{figure}

\begin{figure}
  \resizebox{\hsize}{!}{\includegraphics{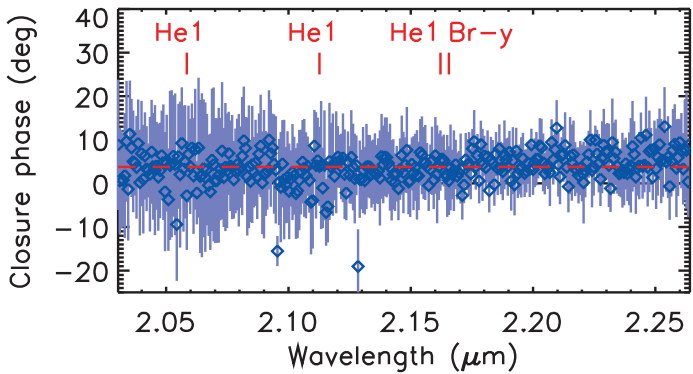}}
  \caption{Calibrated closure phases of \object{Vela X-1} obtained with AMBER. The dashed red line marks the closure phase value in the continuum, and the solid vertical red lines mark the identified spectral features.}\label{fig_amber_clos}
\end{figure}

The calibrated squared visibilities and closure phases measured with AMBER are presented respectively in Figs.~\ref{fig_amber_vis2} and \ref{fig_amber_clos}.

\subsubsection{Spectral analysis}

We analyzed the interferometric observables of \object{Vela~X-1} measured at the wavelengths corresponding to the four spectral lines identified in Sect.~\ref{sec_AMBER_photom}, which are detailed in Table~\ref{tab_AMBER_raies_interf} (see also Figs.~\ref{fig_amber_vis2} and~\ref{fig_amber_clos}). Neither the squared visibilities nor the closure phase signal at these wavelengths are significantly different from the corresponding values in the continuum, estimated from the mean visibilities over the spectral range: the differential signal between the continuum and the lines is at most 3\% and 3\degr{} respectively for the squared visibilities and for the closure phase, but the statistical 1-sigma dispersions of the measurements over all the spectral channels are respectively 3--5\% and 4\degr{} respectively.

As a consequence, this indicates that both the continuum and the lines have the same origin in the stellar wind, and are emitted within very similar formation regions.

\begin{table}
  \caption{Calibrated squared visibilities and closure phase of \object{Vela~X-1} for the four identified spectral lines and in the continuum measured with AMBER.}\label{tab_AMBER_raies_interf}
  \centering 
  \begin{tabular}{c c c c c}
  \hline\hline
  $\lambda$ & $V_1^2$  & $V_2^2$  & $V_3^2$  & $\Phi$\\
    ($\mu$m)  &	(\%) &	(\%)	&   (\%)   &	(\degr)\\
  \hline 
2.058 &$95\pm12$&$80\pm10$&$75\pm9$&$-3\pm13$\\
2.113 &$97\pm12$&$79\pm9$&$71\pm9$&$4\pm10$\\
2.162 &$105\pm12$&$82\pm9$&$70\pm8$&$5\pm7$\\
2.166 &$96\pm11$&$80\pm9$&$74\pm9$&$1\pm8$\\
\hline
Cont. &$98\pm11$&$81\pm9$&$77\pm9$&$4\pm9$\\
  \hline 
  \end{tabular}
\tablefoot{The 2nd to 4th columns give the squared visibilities measured respectively on baselines UT3-UT4, UT1-UT3 and UT1-UT4. The error bars in the continuum corresponds to the averaged uncertainty of the observable over the wavelengths.}
\end{table}

\subsubsection{Parametric modeling with the AMBER data\label{sec_amber_model}}
The values of the interferometric observables in the continuum detailed in Table \ref{tab_AMBER_raies_interf} shows that the system is partially resolved on the two longest baselines.We draw attention from the reader on the fact that only one calibrator has been used for this observation, which can fail to remove some systematic errors in the science measurements. However, the calibrator HD~76304 has been carefully selected in the catalog from \citet{Merand2005}, dedicated to provide reliable calibrator with precise angular diameter estimations for long-baseline optical interferometry. In addition, we checked that the piston level between the acquisitions of \object{Vela X-1} (35.7$~\mu$m rms) and of the calibrator (34.7$~\mu$m then 32.5$~\mu$m) are comparable and does not bias the calibration of the interferometric observables.

As the typical accuracy on the closure phase is 9\degr{} and includes possible systematic errors, the closure phase signal obtained with AMBER does not show any significant departure from the null value (see Fig.~\ref{fig_amber_clos}). As a consequence, we discarded it from the parametric modeling in order to reduce the number of degrees of freedom and improve the precision on the fit parameter. We then used the centro-symmetric model of a uniform disk, which produces a null closure phase signal before the first zero of the visibility function. The diameter of the disk is fitted to the measured squared visibilities. The error bar on the fitted diameter is computed with a variation of 1 from the minimum value of the reduced $\chi^2$, defined as:
\begin{equation}
  \chi^2_\mathrm{red}(D)=\frac{1}{N_\mathrm{free}}\sum_{i=1}^N \left(\frac{y_i-f(x_i,D)}{\sigma_i}\right)^2 ,
\end{equation}
with $\lbrace y_i \rbrace_{i\in[1,N]}$ the measured squared visibilities, $\lbrace \sigma_i \rbrace_{i\in[1,N]}$ the corresponding uncertainties, $\lbrace x_i \rbrace_{i\in[1,N]}$ the corresponding spatial frequencies, $f$ the squared visibility function of a uniform disk, $D$ the disk diameter, and $N_\mathrm{free}$ the number of degrees of freedom.

\begin{figure}
  \resizebox{\hsize}{!}{\includegraphics{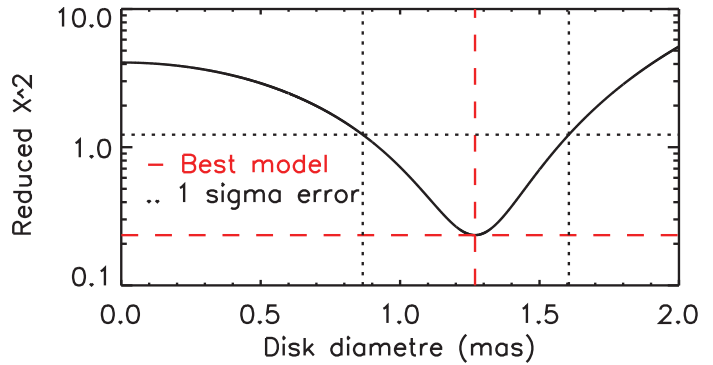}}
  \caption{Reduced $\chi^2$ between the AMBER data and a model of a uniform disk, as a function of the disk diameter. The 1-sigma error bars of the best diameter are computed by a variation of the reduced $\chi^2$ of one (black dotted lines) from the minimum (red dashed lines).}\label{fig_Amber_chi2}
\end{figure}

\begin{figure}
  \resizebox{\hsize}{!}{\includegraphics{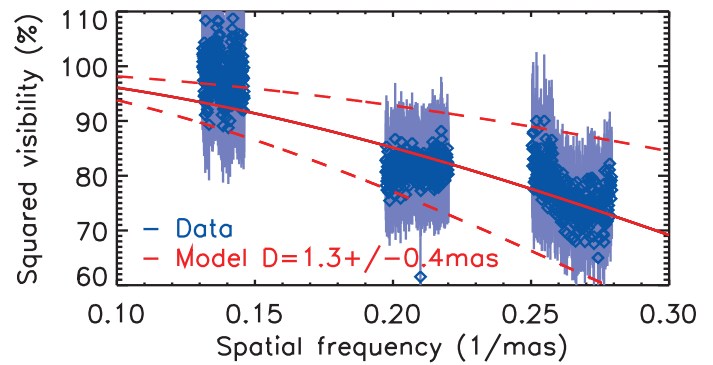}}
  \caption{Squared visibilities of \object{Vela~X-1} measured with AMBER (blue diamonds), and best model of a uniform disk an angular diameter of $1.27 \pm 0.40$~mas (solid red line, error bars in dashed red lines).}\label{fig_Amber_model}
\end{figure}

We found a value of $1.27 \pm 0.40$ mas for the best fit diameter of a uniform disk. The reduced $\chi^2$ as a function of the disk diameter is presented in Fig.~\ref{fig_Amber_chi2}. The squared visibilities of \object{Vela~X-1} measured with AMBER and computed with the best model are shown in Fig.~\ref{fig_Amber_model}. Assuming a distance of 1.9~kpc from Earth \citep{Sadakane1985}, it corresponds to a disk radius of $8 \pm 3~R_\star$.

\subsection{PIONIER interferometric analysis}

\begin{figure}
  \resizebox{\hsize}{!}{\includegraphics{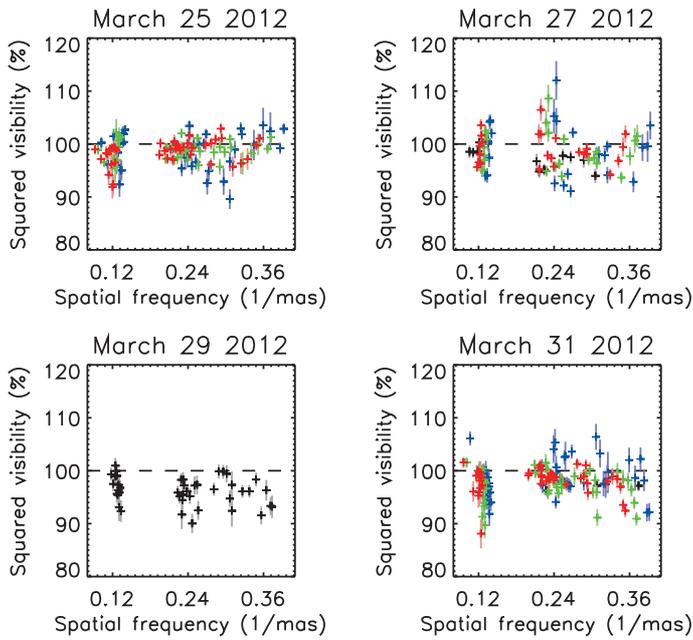}}
  \caption{Calibrated squared visibilities of \object{Vela X-1} obtained with PIONIER for each observing nights. Black symbols stands for broad band measurements, colored symbols stands for low spectral resolution measurements (blue at 1.59~$\mu$m, green at 1.68~$\mu$m, red at 1.76~$\mu$m). The dashed black line marks the squared visibilities of an unresolved target.}\label{fig_pionier_vis2}
\end{figure}

\begin{figure}
  \resizebox{\hsize}{!}{\includegraphics{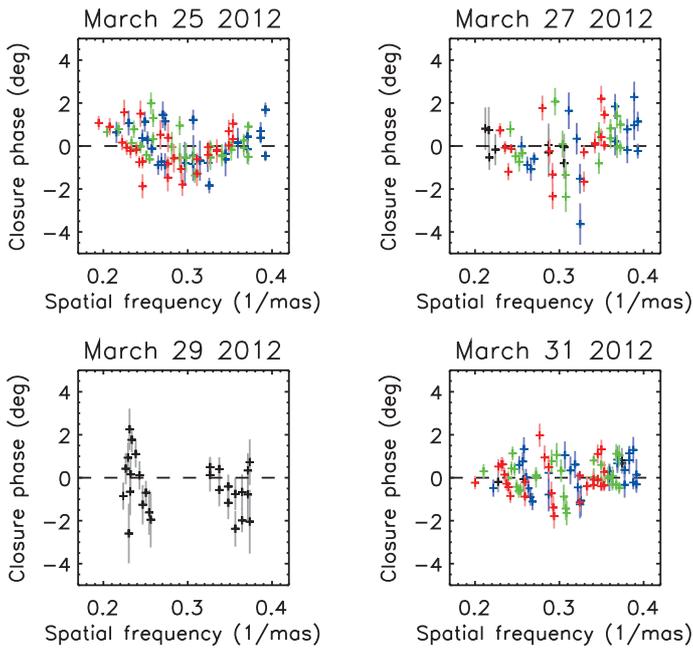}}
  \caption{Calibrated closure phases of \object{Vela X-1} obtained with PIONIER. Same colors as in Fig.~\ref{fig_pionier_vis2}. The dashed black line marks the null closure phase of an unresolved target.}\label{fig_pionier_clos}
\end{figure}

The calibrated squared visibilities and closure phases measured with PIONIER are presented respectively in Figs.~\ref{fig_pionier_vis2} and \ref{fig_pionier_clos}.

\subsubsection{Analysis over an orbital period}

\begin{figure}
  \resizebox{\hsize}{!}{\includegraphics{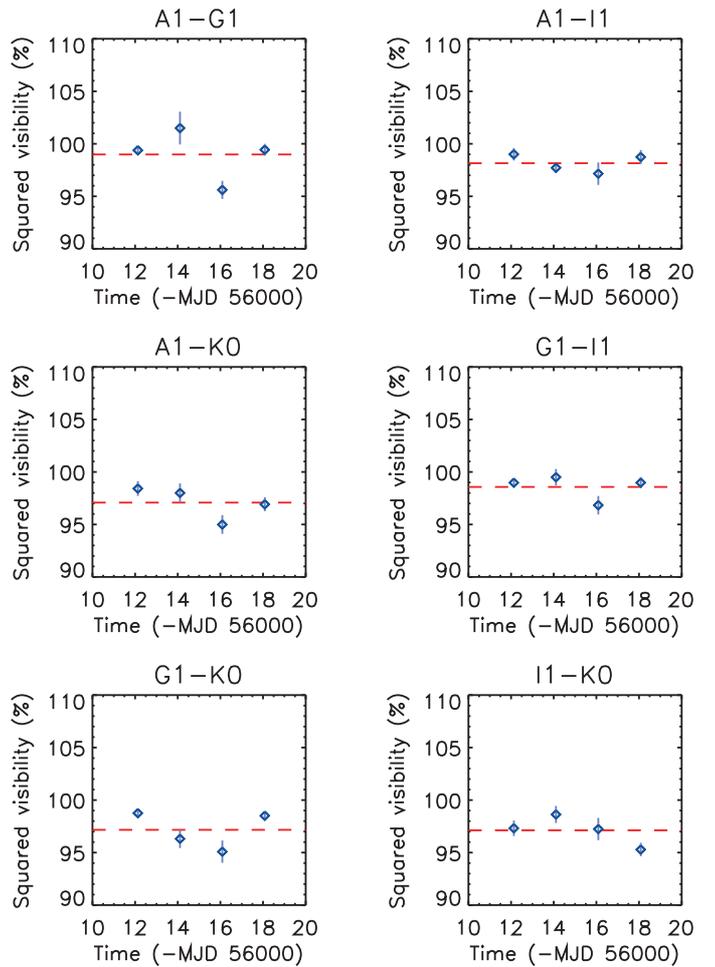}}
  \caption{Squared visibilities of \object{Vela~X-1} measured with PIONIER, averaged for each observing half-night. The error bars show the precision on the average. The red dashed lines mark the mean value over the four nights.}\label{fig_PIONIER_vis2_time}
\end{figure}

\begin{figure}
  \resizebox{\hsize}{!}{\includegraphics{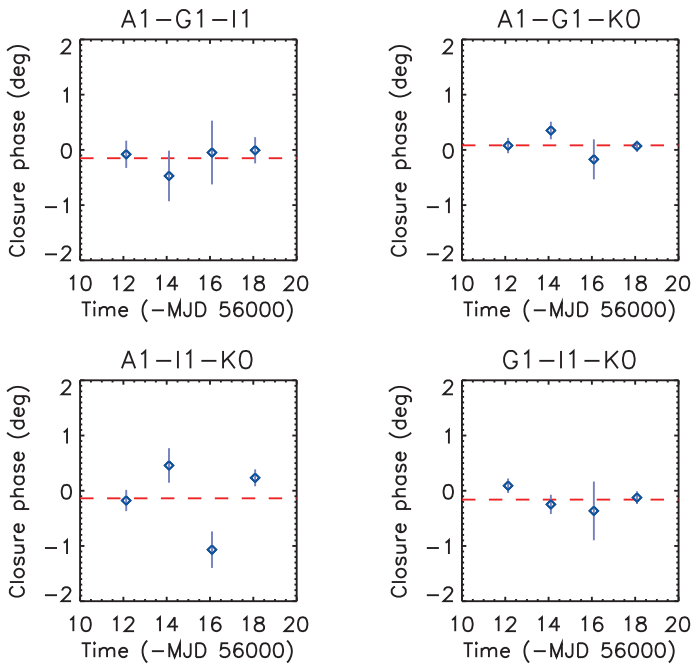}}
  \caption{Closure phases of \object{Vela~X-1} measured with PIONIER, averaged for each observing half-night. As in Fig.~\ref{fig_PIONIER_vis2_time}, the error bars show the precision on the mean of the data. The red dashed lines correspond to the mean value over the four observations.}\label{fig_PIONIER_clot_time}
\end{figure}

As a qualitative analysis, we compared the six squared visibilities and the four closure phases averaged over the half-nights for each observing night. The values are presented respectively in Figs.~\ref{fig_PIONIER_vis2_time} and~\ref{fig_PIONIER_clot_time}. We considered that their variation with the super-synthesis effect provided by Earth rotation along a night is not significant regarding the accuracy of the measurements. This is relevant since the dispersion of the data over a night is similar to their individual uncertainties.

We found that the variation of the interferometric signal over the four nights is marginally significant. The maximum variation is 6\% and 1.5\degr{} respectively for the squared visibilities and the closure phases, which corresponds respectively to 2 and 1.6 times the typical spread of the data over a night. As a consequence, we conclude that the geometry and the symmetry of \object{Vela~X-1} slightly varied with the orbital phase during our observations. We do not further analyze these low-SNR signals for lack of data.

\subsubsection{Parametric modeling with the PIONIER data}

Considering the previous analysis, we adjusted the parametric model of a uniform disk to the entire set of the PIONIER data, regardless of the orbital phase of the system. As the closure phase signals do not show significant departures from the null value (see Fig.~\ref{fig_PIONIER_clot_time}), we chose the centro-symmetric model of a uniform disk, and fit its diameter only to the squared visibility measurements. The precision on the best fit diameter is estimated with a variation of one of the reduced $\chi^2$ from the minimum value, as  in Sect.~\ref{sec_amber_model}.

\begin{figure}
  \resizebox{\hsize}{!}{\includegraphics{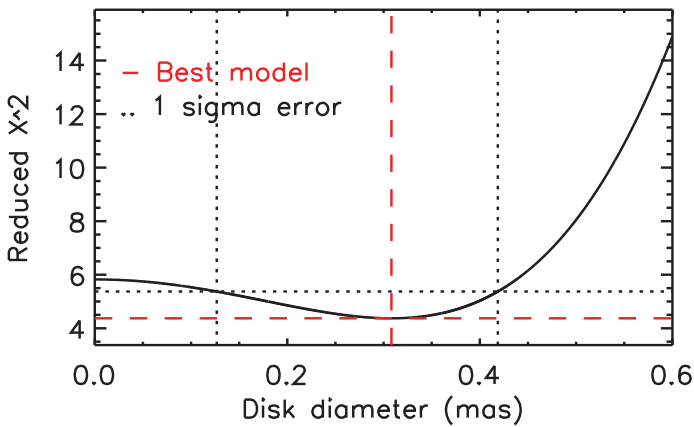}}
  \caption{Reduced $\chi^2$ between the squared visibilities measured with PIONIER and computed with a model of a uniform disk, as a function of the disk diameter. The 1-sigma error bars (black dotted lines) of the best diameter are computed by a variation of the reduced $\chi^2$ of one from the minimum value (red dashed lines).}\label{fig_PIONIER_chi2map}
\end{figure}

\begin{figure}
  \resizebox{\hsize}{!}{\includegraphics{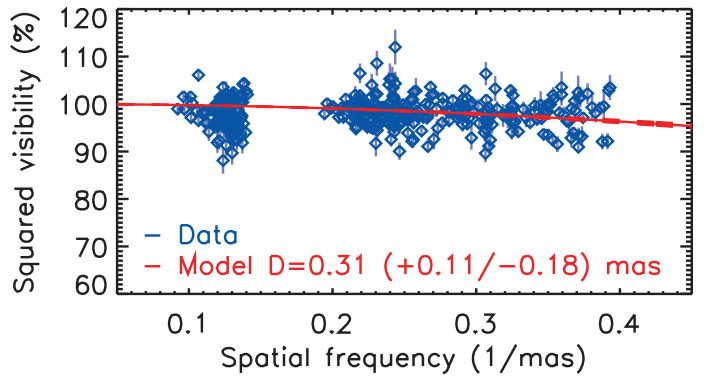}}
  \caption{All squared visibilities of \object{Vela~X-1} measured with PIONIER between March 25 and 31 2012 (blue diamonds), and best model of a uniform disk of diameter $0.31_{-0.18}^{+0.11}$~mas (red line).}\label{fig_PIONIER_fit}
\end{figure}

The reduced $\chi^2$ curve is presented in Fig.~\ref{fig_PIONIER_chi2map}, and the best fit is shown in Fig.~\ref{fig_PIONIER_fit}. We found that the best model corresponds to a disk of diameter $0.31\,_{-0.18}^{+0.11}$~mas, with asymmetric error bars. Assuming a distance of $1.9$~kpc, this corresponds to a structure of radius $2.0\,_{-1.2}^{+0.7}~R_\star$.

\section{Discussion\label{sec_discu}}

During these two campaigns, we observed structures around \object{Vela~X-1} with significantly different sizes, of $8 \pm 3~R_\star$ and $2.0\,_{-1.2}^{+0.7}~R_\star$ in 2010 and 2012 respectively. We discuss here several possible explanations for these different results.

As the observations were carried out in different spectral bands, this difference in the measured diameters can be attributed to a thermal inhomogeneity in the wind. The temperature in hot supergiants indeed varies typically from 10\,000~K at 1--2~$R_\star$ to 5000~K at 5~$R_\star$ \citep{Cidale1998}. The relation between the temperature law in the stellar wind and the size of the envelope observed in the H and K band is however hard to predict without radiative transfer modeling, which is out of the scope of this study. We then compare our estimations to a basic temperature power law. Considering that the effective observing wavelength $\lambda_\mathrm{max}$ reflects the maximum emission of a black body at temperature $T$, the Wien's law provides an estimation of the temperature of the observed material:
\begin{equation}
  \lambda_\mathrm{max}\,T=2.9 \times 10^{-3} \quad \text{m.K}
\end{equation}
Thus, our AMBER observations in the K and H bands can be related to parts of the wind at different temperatures, respectively to a cooler part at 1350~K and a hotter part at 1720~K. Assuming that this temperature gradient can be described with a power law such that $R \propto T^\alpha$, our measurements lead to a power law with $\alpha=-6$, which is significantly different from a black body at thermal equilibrium ($R \propto T^{-2}$, stefan-Boltzman law). 

The second possible explanation for the discrepancy in the measured sizes is that we may have observed a diffuse gaseous shell around the system in 2010 that were no longer present during our observations in 2012. Assuming that the gas propagates at the terminal velocity of the wind, the lifetime of such a shell would be a few days. Such an envelope could be related to the accretion activity of \object{Vela~X-1}, that is known to be highly variable from X-ray observations \citep{Kreykenbohm2008,Furst2010}. Actually, the X-ray luminosity of the compact object irregularly turns off, leading to off-states lasting from a few seconds to one hour and during which the flux drops by a factor ten or more. Analysis of these off-states by \citet{Doroshenko2011} demonstrates that they are due to a drop in the mass accretion rate of the pulsar (whose origin is still debated).
Assuming a drop in the pulsar accretion rate of a factor 20, we estimate the optical density of a shell at 8~$R_\star$ to be 1.3 times more important than the circumstellar matter at 2~$R_\star$ in a regular accretion state of the compact object.
In addition, \citet{Bosch-Ramon2012} showed that in binary systems hosting a massive star and a non-accreting pulsar, the winds of both objects collide and produce a shocked flow extending up to a few times their angular separation, and spiraling with the orbital motion of the system. We could have witnessed such a shocked structure in 2010, produced during an off-state of the pulsar, but spherical instead of spiraling, since seen edge-on. 

The last possibility is that we observed two different components, ie the free-free emission of the stellar wind in the K-band observations, and the supergiant photosphere in the H band observations. Actually, with an emission power spectrum decreasing with the shorter wavelength and a cut-off wavelength in the near infra-red at $\simeq3~\mu$m, the free-free emission is less important in the H band than in the K band, and the wind emission in the H band may be negligible compared to the supergiant photosphere. In addition, by extrapolation of the photosphere diameter of Rigel (B8{Ia}, 237~pc) measured by \citet{Aufdenberg2008}, the photosphere of \object{Vela~X-1} should have a diameter of 0.34~mas, which corresponds to our PIONIER measurement. However, the size of the photosphere of \object{Vela~X-1} reported in the literature \citep{Rawls2011,Quaintrell2003}, corresponding to an angular diameter of 0.16~mas, was computed from radial velocity measurements with a high precision compared to our measurements. As a consequence, we can not safely state if we observed the supergiant photosphere or the stellar wind close to supergiant with PIONIER observations.

\section{Conclusion\label{sec_conc}}

We presented results of near-infrared interferometric observations of \object{Vela~X-1}, carried out at VLTI at two different epochs (March 2010 and March 2012), and in two different spectral bands (respectively in the K and H bands). 

A centro-symmetric structure with a typical radius of $8\pm3$ stellar radii was resolved with the first instrumental setup at $2.2~\mu$m. From the medium spectral resolution of the observations, we deduced that the system presents the same spectral features than a B0.5 {Ib} supergiant, and that the source of the four identified spectral lines is in the same formation region as the continuum. However, despite the better angular resolution provided by the second instrumental setup in the H band, no significant structure was observed in 2012, leading to a centro-symmetric structure of radius $2.0\,_{-1.2}^{+0.7}$ stellar radii at $1.65~\mu$m. This second set sampled the system orbital phase of \object{Vela~X-1} with four observations, but no significant variation was detected in the interferometric observables.

We discussed three possible explanations for this discrepancy in the diameter measurements:
\begin{itemize}
  \item the difference can come from a strong temperature gradient in the supergiant wind, that could explain why the observed structure was much more extended at 2.2~$\mu$m than at 1.65~$\mu$m;
  \item we could have observed a diffuse gaseous shell in 2010 that could have dissolved into the interstellar medium in 2012. Such a structure could have been produced by an off-state in the accretion rate of the pulsar, shocked by the collision between the winds of the supergiant and of the pulsar;
  \item we may have observed directly the supergiant photosphere instead of the stellar wind in our PIONIER observations.
\end{itemize}

To rule out one or the other scenario, further observations should be carried out simultaneously in both near-infrared spectral bands (H and K). In addition, simultaneous (or shortly previous) observations in X-rays would provide informations on the accretion activity of the pulsar, which could be related to the geometry of the stellar wind at few stellar radii of the system. Depending on the results of these observations, radiative transfer modeling should be performed to better understand either how the free-free emission scales in the wind of such a supergiant \citep[as performed by][]{Chesneau2010,Kaufer2012}, or how the intensity distribution observed in 2010 is related to the shocked wind predicted by \citet{Bosch-Ramon2012}.

\begin{acknowledgements}
We are grateful to Olivier Chesneau for his helpful comments on the wind properties of B supergiants.
This research has made use 
of CDS Astronomical Databases SIMBAD and VIZIER, 
of NASA's Astrophysics Data System (ADS), 
of the \texttt{Aspro} and \texttt{SearchCal} services and of the \texttt{AMBER} data reduction package provided by the Jean-Marie Mariotti Center\footnote{Softwares available at http://www.jmmc.fr}. 
The research leading to these results has received funding from the European Community's Seventh Framework Program under Grant Agreement 226604 (Fizeau exchange program).
\end{acknowledgements}

\bibliographystyle{aa}
\bibliography{biblio_HMXB.bib,biblio-gravity-noabb.bib}

\end{document}